# Flame Wrinkles From the Zhdanov-Trubnikov Equation.

Guy Joulin[1#] and Bruno Denet[2*]

[1] Institut P-prime, UPR 3346 CNRS, ENSMA, Université de Poitiers,
1 rue Clément Ader, B.P. 40109, 86961 Futuroscope Cedex, Poitiers, France.

[2] Aix-Marseille Univ., IRPHE, UMR 7342 CNRS,
Technopole de Château-Gombert, 49 rue Joliot-Curie, 13384 Marseille Cedex 13, France.

**Abstract**.

The Zhdanov-Trubnikov equation describing wrinkled premixed flames is studied, using pole-decompositions as starting points. Its one-parameter ($-1 \le c \le 1$) nonlinearity generalizes the Michelson-Sivashinsky equation ($c=0$) to a stronger Darrieus-Landau instability.
The shapes of steady flame crests (or periodic cells) are deduced from Laguerre (or Jacobi) polynomials when $c \approx -1$, which numerical resolutions confirm. Large wrinkles are analysed *via* a pole density: adapting results of Dunkl relates their shapes to the generating function of Meixner-Pollaczek polynomials, which numerical results confirm for $-1 < c \le 0$ (reduced stabilization). Although locally ill-behaved if $c > 0$ (over-stabilization) such analytical solutions can yield accurate flame shapes for $0 \le c \le 0.6$. Open problems are invoked.

*Keywords*: Flame pattern, Nonlinearity, Pole decomposition, Integral equation.

## 1. Introduction.

The flames propagating into premixed gaseous reactants – be they molecular [1] or nuclear [2] – may often be viewed as fronts: their actual thickness $\ell$ based on flat-flame speed $u_L$ and heat diffusivity often is much smaller than the wavelength of their deformations. Being also very subsonic ($u_L \ll$ speed of sound) such combustion fronts border fluids of constant densities: $\rho_u$ (fresh side) or $\rho_b < \rho_u$. As the Atwood number $\mathcal{A} \equiv (\rho_u - \rho_b)/(\rho_u + \rho_b) < 1$ is nonzero, flames are subject to the hydrodynamic Darrieus [3]-Landau [4] (DL) instability at long wavelengths; at shorter scales, variations in local burning speed (relative to reactants) with front mean curvature provide a neutral wavelength $L_{neutral} \sim \ell$ ($L_{neutral} \gg \ell$, though).

Acknowledging that $\mathcal{A} \ll 1$ implies a weak DL instability, Sivashinsky's analysis [5] provided the first systematic weakly-nonlinear description of the local amplitude $\Phi$ of wrinkling: a companion numerical work [6] confirmed that the (Michelson-) Sivashinsky (MS) equation ($c = 0$ in Eq.(1)) correctly captures the slow spontaneous dynamics of flat-on-average flames... if $\mathcal{A} \to 0^+$. Yet in practice $\mathcal{A}$ ranges from 0.2-0.5 (Supernovae) to 0.65-0.85 (chemical fronts), and incorporating higher orders in $\mathcal{A}$ is of interest. The two sub-leading orders just gave the MS equation modified coefficients [7]-[9]. At the next one a new quadratic non-linearity appears in the equation for $\Phi$, but another order only changes coefficients therein [10]. A non-linearity of same type first appeared in $\mathcal{A} = O(1)$ amplitude-expansion derivations of an equation for $\Phi$, where a small flame slope is postulated [11][12].

[#] guy.joulin@lcd.ensma.fr: Corresponding author, Tel(33)(0)549498186, Fax(33)(0)549498176
[*] bruno.denet@irphe.univ-mrs.fr

If a single space coordinate is retained, the Zhdanov-Trubnikov (ZT) type of equation [11] so obtained for slow spontaneous evolutions of front wrinkles has the (rescaled) form:

$$\phi_t + \tfrac{1}{2}\left[\phi_x^2 + c\,\mathcal{H}\{\phi_x\}^2\right] = \nu\phi_{xx} - \mathcal{H}\{\phi_x\}, \qquad (1)$$

where $c = c(\mathcal{A})$, $\phi(t,x) \sim \Phi$ measures the front deformation, and the subscripts denote derivatives with respect to scaled time ($t$) or coordinate ($x$). Provided the reference length is suitably chosen, $\nu > 0$ denotes the neutral-to-actual wavelength ratio of *periodic cells*; for *isolated crests*, $\nu$ could be rescaled to unity by a change of variables but is kept in (1) for comparisons with cells. Curvature effects gave $\nu\phi_{xx}$ whereas the Hilbert non-local transform $\pi\mathcal{H}\{\phi_x\} := \fint_{-\infty}^{+\infty} \phi_x(x')dx'/(x'-x)$ encodes the DL instability: a small $\phi(t,x) \sim \exp(\varpi t + i\kappa x)$ has $\varpi = |\kappa| - \nu\kappa^2$ as growth rate. The stabilizing local nonlinearity in (1) combines geometry [5] and hydrodynamics [7]. The non-local one, absent from the MS equation, mainly is fluid-mechanical [*e.g.* kinetic energy $\sim (\phi_x^2 + \mathcal{H}\{\phi_x\}^2)$]; it may be destabilizing, $c < 0$, or over-stabilizing ($c > 0$), affecting the solutions of (1) in qualitatively different ways (*e.g.* at tips).

Solving the non-linear and non-local Eq.(1) with a general $c$ might help one fit flame shapes from experiments [1], or simulations that use $\mathcal{A} = O(1)$, and is interesting *per se*. The novel solutions of (1), periodic or not, obtained in the present Letter are steps in this direction.

**2. Pole decompositions**.

Just like the MS equation [13], (1) admits a pole decomposition [14]: $\phi_x(t,z)$, the flame slope continued in complex $z := x + iB$ plane, is meromorphic with simple poles $z_k(t)$. All have $-2\nu/(1-c)$ as residue because of a common dominant balance $\phi_x^2 + c\mathcal{H}\{\phi_x\}^2 \approx 2\nu\phi_{xx}$ at $z \approx z_k(t)$. They feature in $N$ conjugate pairs $z_{-k} = \bar{z}_k$ [$\phi(t,z)$ is real when $z$ is] and obey $2N$ coupled differential equations [$dB_k/dt$ is added to the r.h.s. of (3)(6)] for (1) to be satisfied [14]. Attention is presently restricted to steady flame shapes. Only two types thereof are to be considered, $2\pi$ -*periodic cells* and non-periodic *isolated crests*, both of which have their poles aligned along the imaginary axis, $z_k = iB_k$ (mod. $2\pi$ in periodic cases); by convention, $B_k < B_{k+1}$ and $\text{sgn}(B_k) = \text{sgn}(k)$. A single alignment (and flame tip) per cell wavelength is chosen for simplicity, and because such patterns constitute attractors when the unsteady (1) is solved with periodic boundary conditions.

*Periodic cells* with tips centred at $x = 0$ (mod. $2\pi$ ) have $\phi(t,x) = -Vt + \phi(x)$ and:

$$\phi_x(x) = \frac{-\nu}{1-c} \sum_{k=-N}^{N} \cot\left[\tfrac{1}{2}(x - iB_k)\right]. \tag{2}$$

For the "steady" ZT equation to be fulfilled the $B_{|k|\geq 1}$'s must obey [14]

$$\nu \sum_{j=-N, j\neq k}^{j=N} \frac{1 - c\varepsilon_k\varepsilon_j}{1-c} \coth\left[\tfrac{1}{2}(B_k - B_j)\right] = f\varepsilon_k, \tag{3}$$

$$f := 1 + 2\nu Nc/(1-c), \tag{4}$$

where $\varepsilon_k := \text{sgn}(B_k)$ originates from $\mathcal{H}\{\cot(\tfrac{1}{2}(x - iB_k))\} = -1 - i\varepsilon_k \cot(\tfrac{1}{2}(x - iB_k))$. The number $N$ of pole pairs in (2) is constrained by the inequality $\coth(\tfrac{1}{2}(B_N - B_{j\neq N})) \geq 1$ to be less than $N_{opt}(\nu) := \lfloor \tfrac{1}{2}(1 + \tfrac{1}{\nu}) \rfloor$ ($\lfloor . \rfloor$=integer part), but is otherwise arbitrary. The speed $V = (1+c)\tfrac{1}{2\pi}\int_0^\pi \phi_x^2 dx$ is found [14] from (2)(3) to be $V = 2N\nu(1 - N\nu)/(1-c)$, whereby $c^2 \leq 1$.

Centred *Isolated crests* are limiting cases of cells ($z \sim N\nu \ll 1$), have $\mathcal{H}\{\phi_x\} \sim \nu^2 N^2 / x^2$ $\sim \phi_x^2$ at $|x|/\nu N \to \infty$, and $V = 0$. For any $N \geq 1$ they obey 'local' versions of (2)-(4), *viz*.:

$$\phi_x(x) = \frac{-\nu}{1-c} \sum_{k=-N}^{N} \frac{2}{x - iB_k}, \tag{5}$$

$$\nu \sum_{j=-N, j\neq k}^{j=N} \frac{1 - c\varepsilon_k\varepsilon_j}{1-c} \frac{2}{B_k - B_j} = \varepsilon_k. \tag{6}$$

As when $c = 0$ [13], the only known exact result is then obtained by summing (6) times-$B_k$ over $k \geq 1$, $0 < B_1 + ... + B_N = \nu N(2N/(1-c) - 1)$, whereby $c < 1$.

### 3. Isolated crests.

For reasons that shall gradually emerge, isolated crests with $c < 0$ are considered first. The simplest of them has $N = 1$. Equation (6) then gives $B_1 / \nu = (1+c)/(1-c)$ , with three consequences: (i) $B_1 > 0$ needs $c^2 < 1$, confirming this range as the physical one; (ii) $B_1/\nu$ increases with $c$; (iii) $B_1/\nu \to 0$ as $c \to -1$, suggesting this is a limit to envisage first.

**3.1: The $c \to -1$ limit.**

A small $1+c$ means that the $B_{k>0}$ only little feel their "conjugates" $B_{k<0}$ ...unless very close; yet the exception can only concern $B_{\pm 1}$, due to the repulsion between $B_k$'s of like signs. One then anticipates that $B_{\pm 1}/\nu$ will be $O(1+c) << 1$, just like when $N=1$, the other poles staying at finite distances. Under this working assumption (6) simplifies considerably for $k = 2,...,N$:

$$2 \sum_{j=2, j\neq k}^{j=N} \frac{\nu}{B_k - B_j} + \frac{\nu(1+\alpha)}{B_k} \approx 1, \ \alpha = 1. \tag{7}$$

According to a classical work of Stieltjes [15] such leading order $B_{k\geq 2}/\nu := \eta_k$ therefore are zeros of the associated Laguerre polynomial $\mathcal{L}_{N-1}^{(\alpha=1)}(\eta)$. Equation (6) also simplifies for $k=1$:

$$\frac{(1+c)\nu}{2B_1} \approx 1 + 2\sum_{j=2}^{N} \frac{\nu}{B_j}. \tag{8}$$

Since $1/\eta_2 + ... + 1/\eta_N = \frac{1}{2}(N-1)$ by (7), $B_1/\nu \approx (1+c)/2N$: as anticipated $B_1/\nu << \eta_2$. On integration (5) then yields the limiting flame shape $\phi(x)$ for $c \to -1$, uniformly in $x$ and $N$, in the Hopf-Cole-like form:

$$\phi = \phi^+(x) + \phi^-(x) + const., \tag{9}$$

$$(1-c)\phi^\pm(z) = -2\nu \ln\left[\psi^\pm(z)\right], \tag{10}$$

where $\nu\psi^\pm(z) = (z \mp iB_1)\mathcal{L}_{N-1}^{(1)}(\pm z/i\nu)$. Numerical resolutions of (6) confirm this: for $N=10$, $1/\nu = 1/20.5$, $c = 0.999$ they give $B_1 = 2.445 \ 10^{-6}$, $B_2 = 0.01799$, $B_6 = 0.3525$, $B_{10} = 1.3722$, while the values predicted from (7)(8) are $2.441 \ 10^{-6}$, $0.01796$, $0.3523$, $1.3716$, respectively.

The above steady $\phi(x)$ has a 2-scale structure, $x/\nu \sim 1$ and $x/\nu \sim (1+c)/N << 1$. This has counterparts for unsteady shapes $\phi(t,z) = \phi^+(t,z) + \phi^-(t,z)$, with $\phi^-(t,z) = \overline{\phi^+(t,\overline{z})}$ as in (9), and $\phi_z^+$ analytic in $\Im(z) \leq 0$ whereby $\mathcal{H}\{\phi_x^\pm\} = \mp i\phi_x^\pm$. Insofar as their lowest poles $iB_{\pm 1}(t)$ have $NB_{\pm 1}(t)/\nu >> (1+c)$, $(\phi^\pm)_{c\to -1}$ follow *uncoupled* Burgers-type (thus linearisable) equations

$$\phi_t^\pm + \phi_z^{\pm 2} \mp i\phi_z^\pm = \nu\phi_{zz}^\pm, \tag{11}$$

since well separated poles $iB_k(t)$ with $k$'s of unlike signs do not interact if $c \to -1$. Those of $\phi^+$ (or $\phi^-$) globally drift towards $\Im(z) = 0$ while interacting among themselves, until $NB_{\pm 1}(t) \sim \nu(1+c) \ll 1$. From then on, $F^\pm := \phi^\pm + \nu \ln(z \mp iB_1(t))$ obey another linearisable equation

$$F_t^\pm + F_z^{\pm 2} \mp iF_z^\pm - \tfrac{2\nu}{z}(F_z^\pm - F_z^\pm(t,0)) = \nu F_{zz}^\pm \qquad (12)$$

at all $|z| \gg B_1(t)$ scales. The near-real $B_{\pm 1}(t)$ induce the $\tfrac{-2\nu}{z}$ "convection" term in (12), and soon get themselves slaved to $F_z^+(t,0) = -F_z^-(t,0)$, $(1+c)\nu/2B_{\pm 1}(t) \approx \pm 1 - 2F_z^\pm(t,0)$ (see (8)). The front shape is then $\phi(t,x) \approx -\nu \ln\left[(x^2 + B_1(t)^2)|F^+(t,x)|^2\right]$; it has a spike on top of the smoother pattern governed by (12) that *in fine* monitors its varying height and width.

**3.2: Large crests**.

Steady poles get packed when $N \to \infty$, $B_{k+1} - B_k \ll B_N$; it is then appropriate to define such a density $\rho(B) = \rho(-B)$ that $\rho(B)dB$ measures the number of poles with 'altitudes' in $(B, B+dB)$ [13]. In this continuous approximation of (6) the pole density pertaining to isolated crests obeys the singular integral equation:

$$\fint \frac{[1 - c\varepsilon(B)\varepsilon(B')]2\nu\rho(B')dB'}{(1-c)(B-B')} = \varepsilon(B), \qquad (13)$$

with $\varepsilon(B) := \mathrm{sgn}(B)$. The principal-part prescription of this integral reflects the constraint $j \neq k$ in (6), and its unspecified integration range is unknown; as when $c = 0$ one anticipates that $\rho(|B| > B_{max}) = 0$ for some $B_{max} > 0$.

When $c = 0$ Eq.(13) was solved [16] by Fourier series methods in terms of an angle $-\pi/2 \leq \theta \leq \pi/2$ defined by $B/B_{max} = \sin(\theta)$. The solution $\rho_{c=0}(B)$ reads:

$$2\nu\pi\rho_{c=0} = w(\theta), \qquad (14)$$

$$\pi w(\theta) := \ln(\cot^2(\tfrac{1}{2}\theta)) \qquad (15)$$

The normalization $\int_0^{B_{max}} \rho_{c=0}(B)dB = N$ fixed $B_{max,c=0}$ to be $2\pi N\nu$ and contour evaluation of

$$\phi_x(x) = \frac{-2\nu}{(1-c)} \int_{-B_{max}}^{B_{max}} \frac{\rho(B)dB}{x - iB} \qquad (16)$$

with $c = 0$ gave the front slope [16], see below (22). A noteworthy point is as follows: although this density has $-B_{max} < B < B_{max}$ as a support, it may be viewed as the restriction to the physical range $-\pi/2 \leq \theta \leq \pi/2$ of a $2\pi$-*periodic* function $\tilde{\rho}(\theta)$ defined for *all* real $\theta$ and vanishing at all odd multiples of $\pi/2$: this odd continuous function of $w(\theta) = w(-\theta)$ may also be assumed to satisfy:

$$\tilde{\rho}(\theta + 2\pi) = \tilde{\rho}(\theta) = \tilde{\rho}(-\theta) = -\tilde{\rho}(\pi - \theta). \tag{17}$$

We anticipate that such symmetries still hold when $c \neq 0$. Trigonometric identities [*e.g.* $2\cos\theta'/(\sin\theta' - \sin\theta) \equiv \cot(\frac{\theta'-\theta}{2}) - \tan(\frac{\theta'+\theta}{2})$] then indicate that (13) will be solved when one finds a $\tilde{\rho}(\theta)$ obeying (17) and

$$\tilde{\varepsilon}(\theta)\mathcal{H}\{\tilde{\rho}(\theta)\} - c\mathcal{H}\{\tilde{\varepsilon}(\theta)\tilde{\rho}(\theta)\} = -(1-c)/2\pi\nu, \tag{18}$$

where $\tilde{\varepsilon}(\theta) := \text{sgn}(\sin(\theta))$ and the (periodic-) Hilbert transform $\mathcal{H}\{.\}$ now refers to $\theta$.

Not being of the convolution type if $c \neq 0$ (18) is no longer simply amenable to a Fourier method. One can fortunately invoke results by Dunkl [17], who exhibited a sequence of one-variable real polynomials $p_j(\omega)$ that obey "ladder" relations ([17], p.149, Theorem 1) once evaluated at $\omega = w(\theta)$:

$$\mathcal{H}\{\tilde{\varepsilon}(\theta)p_j(w(\theta))\} = +p_{j+1}(w(\theta)), \ j = 0,1,\ldots, \tag{19}$$

$$\tilde{\varepsilon}(\theta)\mathcal{H}\{p_j(w(\theta))\} = -p_{j-1}(w(\theta)), \ j = 1,2,\ldots. \tag{20}$$

Direct substitution in (18) and a term-by-term use of (19)(20) show that $2\pi\nu\tilde{\rho}(\theta)/(1-c) = \sum_{n=0}^{\infty}(-c)^n p_{2n+1}(w(\theta))$ satisfies (18) if $p_0(.) = 1$. The $p_j(\omega)$'s so identified are the (non-monic) symmetric Meixner-Pollaczek polynomials $P_j^{(\lambda=1/2)}(\omega)$, with $(-1)^j$ as parity and $\sum_{j=0}^{\infty} r^j p_j(\omega)$ $:= G(\omega, r) = (1+r^2)^{-1/2} e^{\omega \arctan(r)}$ as generating function [15, 17].

The sum $\sum_{n=0}^{\infty}(r^2)^n p_{2n+1}(\omega)$ is thus $\frac{1}{2r}(G(\omega,r) - G(-\omega,r)) = \frac{(1+r^2)^{-1/2}}{r}\sinh(\omega\arctan(r))$. With $r^2 = -c$ and $\omega = w(\theta)$ this yields $2\pi\nu\tilde{\rho}(\theta)/(1-c)$ in closed form, to produce:

$$2\nu\pi\rho(B) = \frac{(1-c)^{1/2}}{\sqrt{-c}}\sinh\left[w(\theta)\arctan(\sqrt{-c})\right] \tag{21}$$

if specialized to the physical range; as $w(\theta)$ itself (21) obeys (17), and resumes (14) for $c = 0$.

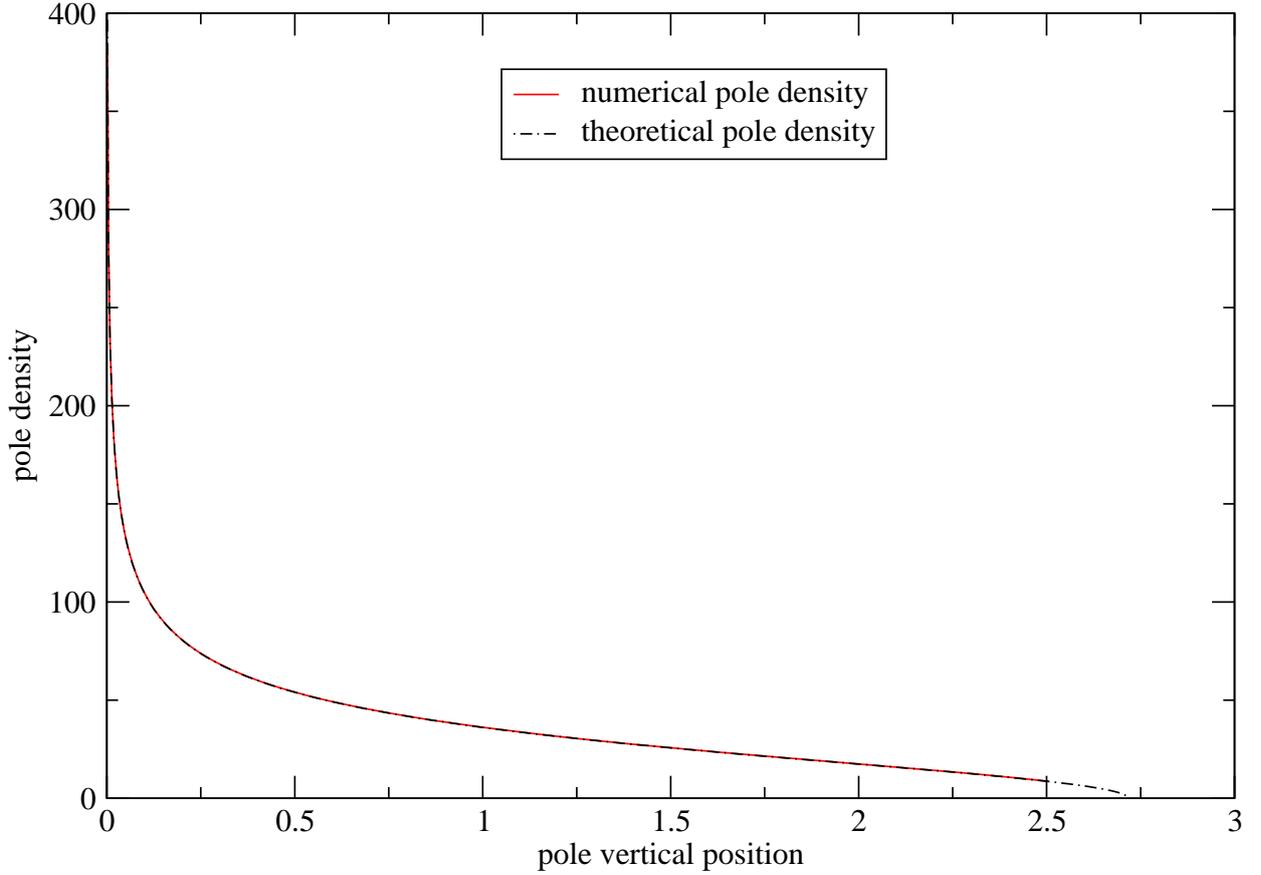

**Figure 1**: Numerical density [defined by $\rho_{num}(\frac{1}{2}(B_k + B_{k-1})) := 1/(B_k - B_{k-1})$ with $B_k$'s for crests from (6)] *vs.* pole altitude $B$, and predicted density (21); $N = 100$, $\nu = 1/199.5$, $c = -0.25$.

Normalisation requires $B_{max} \int_0^{\pi/2} \rho(\theta)\cos(\theta)d\theta = N$ : contour integrations[1] yield $B_{max} = 2\nu N \sin(\pi\mu)/\mu$, $0 \leq \cos(\pi\mu) := (1+c)/(1-c) \leq 1$. This and $\theta := \arcsin(B/B_{max})$ complete the determination of $\rho(B)$ in (21) . For $c = -1$, $2\pi\nu\rho$ acquires the Marchenko-Pastur form $(4\pi\nu/|B|-1)^{1/2}$ pertaining to $\mathcal{L}_{N-1\to\infty}^{(1)}(|B|/\nu)$ [18], as expected (Sec. **3.1**) .

---

[1] This integral reads $\int_0^1 h(t)dt$, $t := \tan(\frac{1}{2}\theta)$, or $\frac{1}{2}\int_0^\infty h(t)dt$ as $h(1/t) \equiv t^2 h(t)$. Integrations by parts then along the contour $\{\Im(u) = 0\} \cup \{\Im(u) = 2\pi\}$ in the complex plane of $u = \ln(t^2)$, and residue evaluations, give $B_{max}$.

The flame slope, Eq.(16), is expressible in terms of $\sinh(\xi) := x/B_{max}$ as:

$$\phi_x(x) = \frac{\text{sgn}(-x)}{2\tan(\tfrac{1}{2}\pi\mu)}\left[|\coth(\tfrac{1}{2}\xi)|^\mu - |\tanh(\tfrac{1}{2}\xi)|^\mu\right], \tag{22}$$

which when $c = 0 = \mu$ resumes the profile $\pi\phi_x(x) = \text{sgn}(-x)\ln\left[\coth^2(\tfrac{1}{2}\xi)\right]$ found in [16]. Figures 1-3 compare (21)(22), and ensuing flame shapes $\phi(x)$, with what Eqs. (6)(5) give if $N \gg 1$, respectively: smaller $c < 0$ (weaker non-linear stabilization) give sharper crest *tips*.

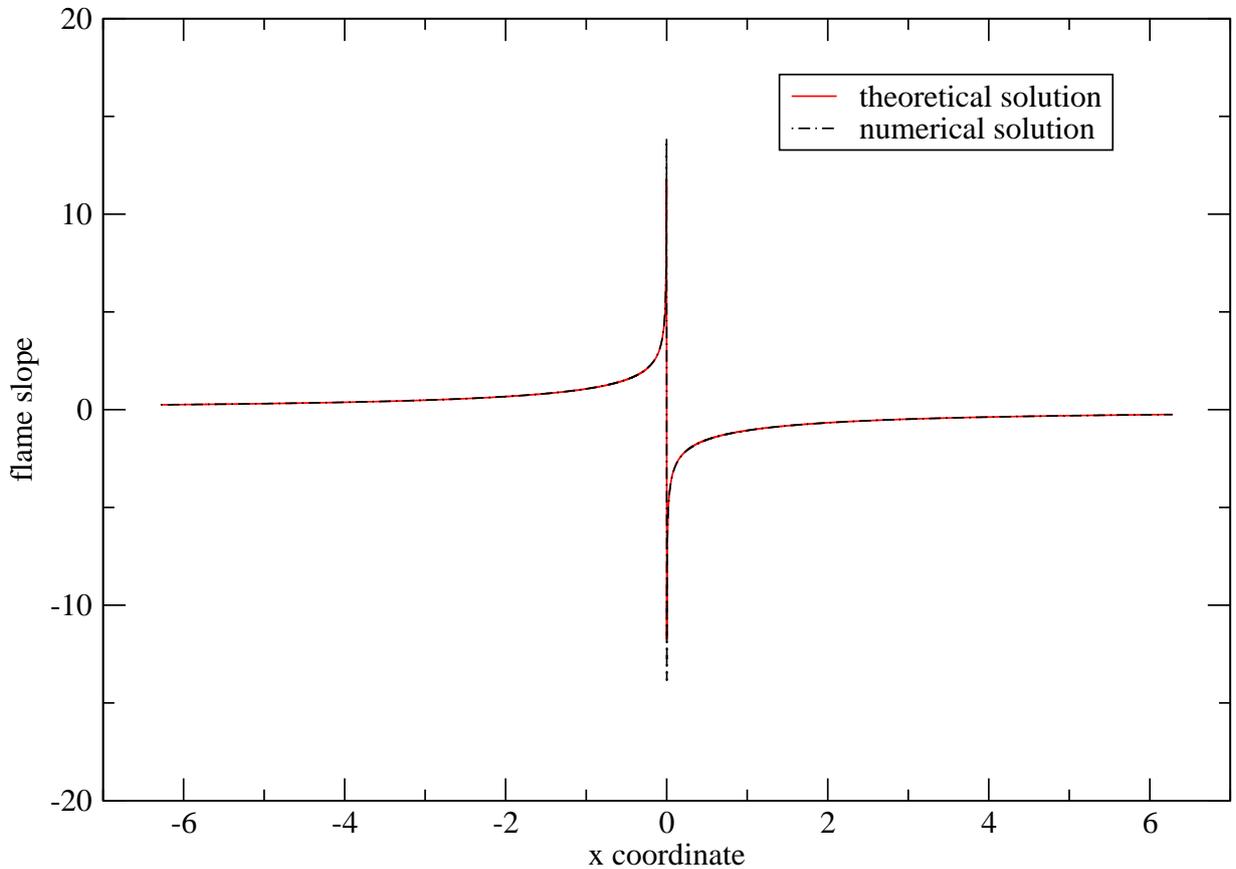

**Figure 2**: Numerical flame-crest slope $\phi_x(x)$ [from (5)(6)] *vs.* abscissa $x$, compared with analytical prediction (22); $N = 100$, $\nu = 1/199.5$, $c = -0.25$.

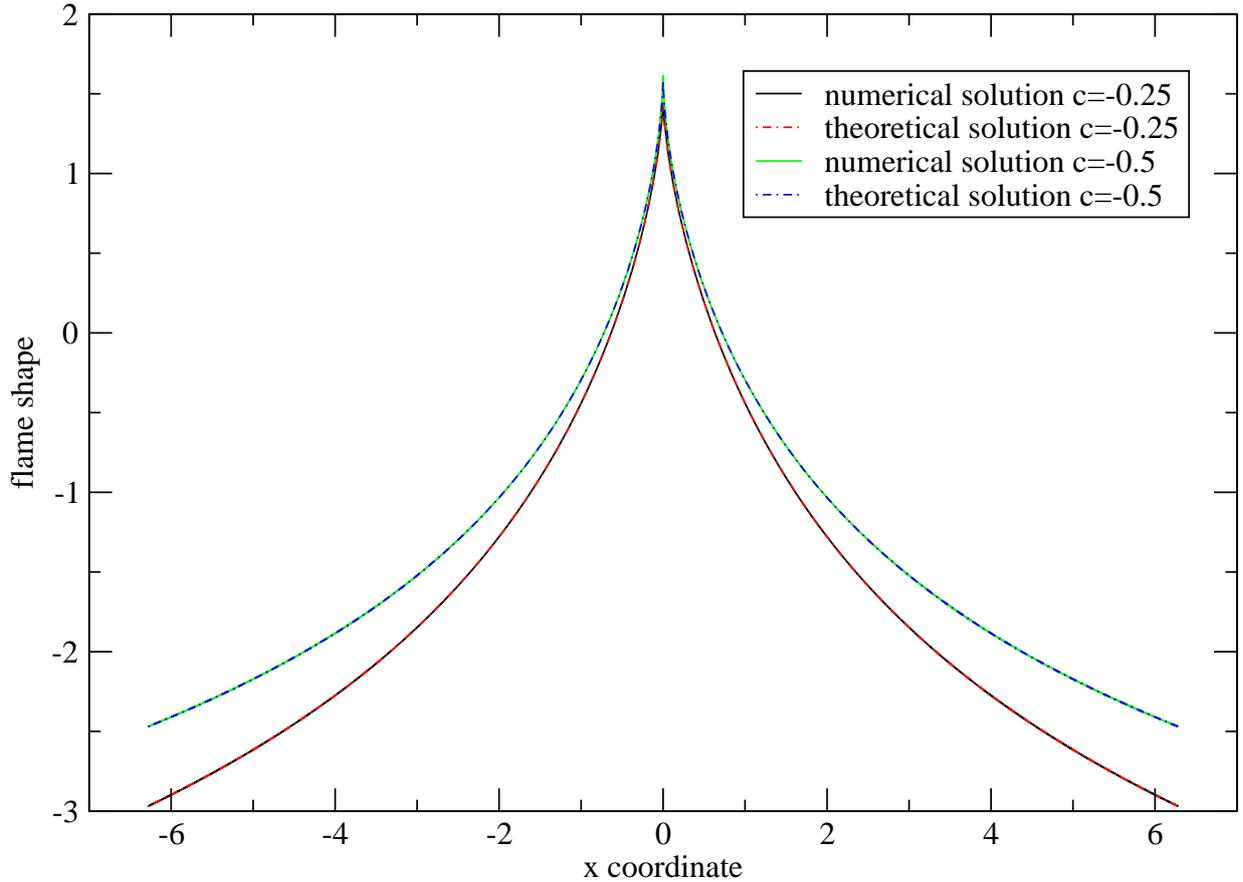

**Figure 3**: Numerical flame-crest shapes $\phi(x)$ [from (5)(6)] *vs.* abscissa $x$ [same parameters as in Fig.1], and predictions from numerical quadrature of Eq.(22).

## 4. Periodic cells.

To analyse $2\pi$-periodic patterns one again starts with $N=1$, in which case (3)(4) give $\tanh(B_1) = \nu(1+c)/(1-c)f$. One checks that $0 < B_1 < \infty$ needs $c^2 < 1$ and $1 = N \leq \frac{1}{2}(1+\frac{1}{\nu})$, *i.e.* $\nu \leq 1$: cells of finite amplitude exist *iff* unstable modes fit in the box $|x| < \pi$.

### 4.1: The $c \to -1$ limit.

$B_1 > 0$ again is well below $B_2, ..., B_N$ if $0 < 1+c = o(1)$. Equations (3)(4) specialized to the latter poles again simplify, especially when expressed in terms of $C_k := \coth(\frac{1}{2}B_k) > 1$:

$$\sum_{j=2, j \neq k}^{j=N} \frac{\nu(1-C_k^2)}{(C_k - C_j)} + \nu(N-1)C_k - 1 - \frac{2N\nu c}{1-c} \approx 0. \qquad (23)$$

Another result of Stieltjes [15] says that such $C_k$'s coincide with the zeros of a Jacobi polynomial $P_{N-1}^{(\alpha_1',\alpha_1'')}(C)$ in $C := \coth(\tfrac{1}{2}B)$, yet with indices $\alpha_1' = 1/\nu - 2N$ and $\alpha_1'' = -1/\nu < 0$ that lead to nonstandard polynomials ($\alpha_1'' < -1$). The identity $P_{N-1}^{(\alpha_1',\alpha_1'')}(X_1)/(1+X_1)^{N-1}$ $\sim P_{N-1}^{(\alpha_2',\alpha_2'')}(X_2)$, with $X_2 = (3-X_1)/(1+X_1)$, $\alpha_2' = \alpha_1'$, $\alpha_2'' = 1 - 2N - \alpha_1' - \alpha_1''$ [19], and another identity changing the first index only, fortunately help one access the $C_{k\geq 2} > 1$: they obey the more manageable $P_{N-1}^{(\alpha',\alpha'')}((C-3)/(C+1)) = 0$, where $\alpha' = 1$ and $\nu\alpha'' = 1 - 2N\nu \geq 0$. Equation (3) also simplifies for $k=1$, giving $B_1 \approx \nu(1+c)/2N(1-N\nu) \ll B_2$ once $C_2 + ... + C_N = 1/\nu - N$ (by (23)) is used.

The flame shape then obeys (9)(10) with $\psi^\pm(z) = \sin(\tfrac{1}{2}(z \mp iB_1))P_{N-1}^{(\alpha',\alpha'')}(2e^{\pm iz} - 1)$, since $(C-3)/(C+1) \equiv 2e^B - 1$. The resulting wrinkle amplitude $\max(\phi) - \min(\phi)$ diverges like $-2\nu\ln(1+c) + const.$ when $c \to -1$ at fixed $\nu$ and $\nu N \sim 1$, because $-\nu\ln\{|\sin(\tfrac{1}{2}(x - iB_1))|^2\}$ endows the front with spikes centred at $x = 0 \pmod{2\pi}$, each of $O(B_1) \sim \nu(1+c)/N$ width.

**4.2: Large cells**.

In the large-$N$ limit the neutral-to-actual wavelength ratio $\nu$ must be small, for compatibility with $N \leq N_{opt}(\nu) \approx \tfrac{1}{2\nu}$. With this proviso, the pole-density $\rho(B)$ now obeys:

$$\oint \frac{[1 - c\varepsilon(B)\varepsilon(B')]\nu\rho(B')dB'}{(1-c)\tanh(\tfrac{1}{2}(B-B'))} = f\varepsilon(B). \qquad (24)$$

The identity $\coth(\tfrac{1}{2}(B-B')) \equiv (1-T'^2)/(T-T') - T'$, with $T := \tanh(\tfrac{1}{2}B)$ and similarly for $T'$, helps one solve this formidable-looking equation. Trading $B$ for $T$ and exploiting the parities of $\rho$ and $\varepsilon(T) = \varepsilon(B)$ indeed produces:

$$\oint \frac{[1 - c\varepsilon(T)\varepsilon(T')]2\nu\rho(T')dT'}{(1-c)(T-T')F} = \varepsilon(T), \qquad (25)$$

$$F := 1 + \frac{2\nu Nc}{1-c} - c\int \frac{2\nu\rho(T')T'\varepsilon(T')dT'}{(1-c)(1-T'^2)}, \qquad (26)$$

which formally coincides with (13), except for the provisionally unknown constant $F \neq 1$. One can nevertheless transpose the result on isolated crests as:

$$2\nu\pi\rho(B)/F = (1-c)^{1/2} \sinh\left[w(\theta)\arctan(\sqrt{-c})\right]/\sqrt{-c}, \qquad (27)$$

where now $\tanh(\tfrac{1}{2}B) := \sin(\theta)\tanh(\tfrac{1}{2}B_{max})$. The integrals $\int_0^{B_{max}} \rho(B)dB = N$ and in definition (26) are amenable to contour integrations [again in $\ln(\tan^2(\tfrac{1}{2}\theta))$-plane, similar to footnote[1]], to ultimately yields all the ingredients needed to characterize $\rho(B)$ in (27):

$$\tanh(\tfrac{1}{2}B_{max}) = \sin(\beta), \; F = \frac{1-c(1-2N\nu)}{\cos(\mu\beta)(1-c)}, \qquad (28)$$

$$\tan(\mu\beta) := \frac{2\nu N\sqrt{-c}}{1-c(1-2N\nu)}, \; \tan(\tfrac{1}{2}\pi\mu) := \sqrt{-c}. \qquad (29)$$

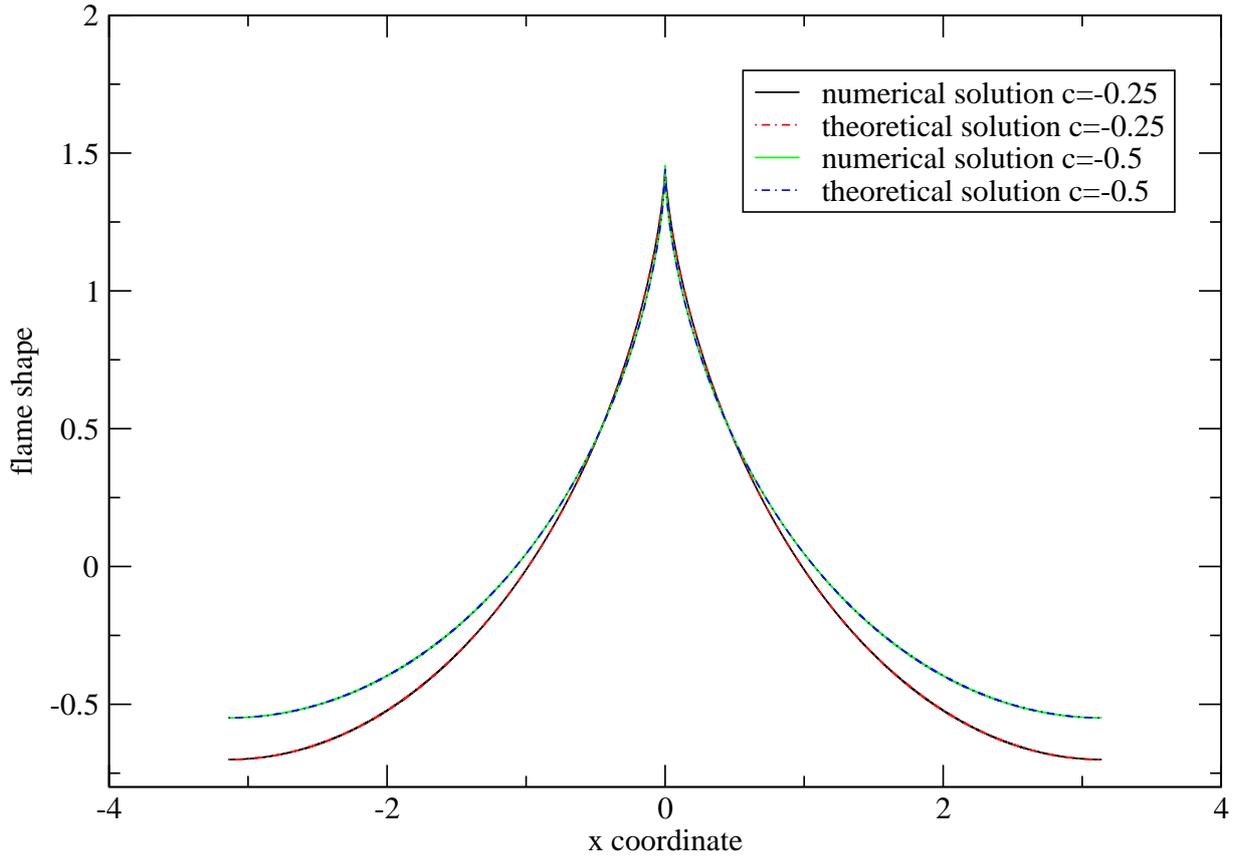

**Figure 4**: Numerical flame-cell shapes $\phi(x)$ [from (2)(3)] *vs.* abscissa $x$ [same parameters as in Fig.1], and predictions from quadrature of Eq.(30).

One checks that: (i) $N\nu \ll 1$ gives $B_{\max}/2 \approx \beta \ll 1$, $F \approx 1$ and $B_{\max}$ near to the same $2\nu N \sin(\mu\pi)/\mu$ as for isolated crests; (ii) $B_{\max} = \infty$ if $2\nu N = 1 \approx 2\nu N_{opt}$ ; (iii) At $c = -1$, $2\pi\nu\rho = (1 - N\nu)\left[\tanh(\tfrac{1}{2}B_{\max})/\tanh(\tfrac{1}{2}B) - 1\right]^{1/2}$. The flame slope resembles (22):

$$\phi_x(x) = \frac{\operatorname{sgn}(-x)F}{2\tan(\tfrac{1}{2}\pi\mu)}\left[|\coth(\tfrac{1}{2}\xi)|^{\mu} - |\tanh(\tfrac{1}{2}\xi)|^{\mu}\right], \tag{30}$$

yet with $\sinh(\xi) = \tan(\tfrac{1}{2}x)/\tanh(\tfrac{1}{2}B_{\max})$. When $2\nu N = 1$ [ $\beta = \tfrac{\pi}{2}$, $B_{\max} = \infty$, $\tanh(\tfrac{1}{2}\xi) = \tan(\tfrac{1}{4}x)$, $F = \cos(\tfrac{1}{2}\pi\mu)$ ] the flame shape is expressible *via* hypergeometric functions $_2F_1[.,.;.;.]$:

$$\phi(x) = 2(\Upsilon(x,\mu) - \Upsilon(x,-\mu))\cos(\tfrac{1}{2}\pi\mu)/\tan(\tfrac{1}{2}\pi\mu), \tag{31}$$

where $(1+\mu)\Upsilon(x,\mu) := {_2F_1}\left[1, \tfrac{1+\mu}{2}; 1+\tfrac{1+\mu}{2}; -\tan^2(\tfrac{1}{4}x)\right]\tan^{1+\mu}(\tfrac{1}{4}|x|)$.

Numerical resolutions of (3) with $N \gg 1$ confirm the above findings, see Fig. 4: although $c < 0$ implies sharp tips, $[\phi] := -\int_{0^+}^{\pi}\phi_x dx$ *decreases* as $|c|$ grows. And $[\phi]$ remains finite at $c = -1$: continuous densities cannot capture the sharp spikes caused by $NB_1 = o(\nu)$ (Sec. **4.1**).

## 5. Positive $c$ 's.

The procedure leading to (21)(27) formally applies to $0 < c < 1$ (one may still use the generating function $G(\omega, r)$ with $r^2 = -c$, since $|r| < 1$ [17]): $\sqrt{-c} = ic^{1/2}$ just gets imaginary, as does the exponent $\mu = im$ now related to $c$ by $1 < (1+c)/(1-c) = \cosh(m\pi)$. The pole density (21) previously tailored for isolated crests with $c < 0$ simply becomes:

$$2\nu\pi\rho(B) = (1-c)^{1/2}\sin\left[w(\theta)\operatorname{atanh}(c^{1/2})\right]/c^{1/2}. \tag{32}$$

This $\rho(B)$ wildly oscillates as $|B|$ decreases below $B_{\max}\sin(\theta_0)$, $\tan(\tfrac{1}{2}\theta_0) := e^{-\pi/m}$, as does $\phi_x(x) = \operatorname{sgn}(-x)\coth(\tfrac{1}{2}\pi m)\sin\left[m\ln(\coth(\tfrac{1}{2}|\xi|))\right]$ near $x = 0$: although an exact solution of (13) the above $\rho(B)$ is physically *spurious*, because acceptable pole-densities are nonnegative.

Still, when $c > 0$ is small enough, numerical resolutions of (3) for large isolated crests reveal that the above $\rho(B)$ accurately fits the numerical density defined by $\rho_{num}(\tfrac{1}{2}(B_k + B_{k-1}))$ $:= 1/(B_k - B_{k-1})$, if (32) uses $B_{\max} = 2\nu N\sinh(m\pi)/m$ as $\int_0^{B_{\max}}\rho(B)dB = N$ would imply; the resulting $\phi_x(x)$ very accurately fits the numerical slopes, except *very* close to $x = 0$, Fig.5.

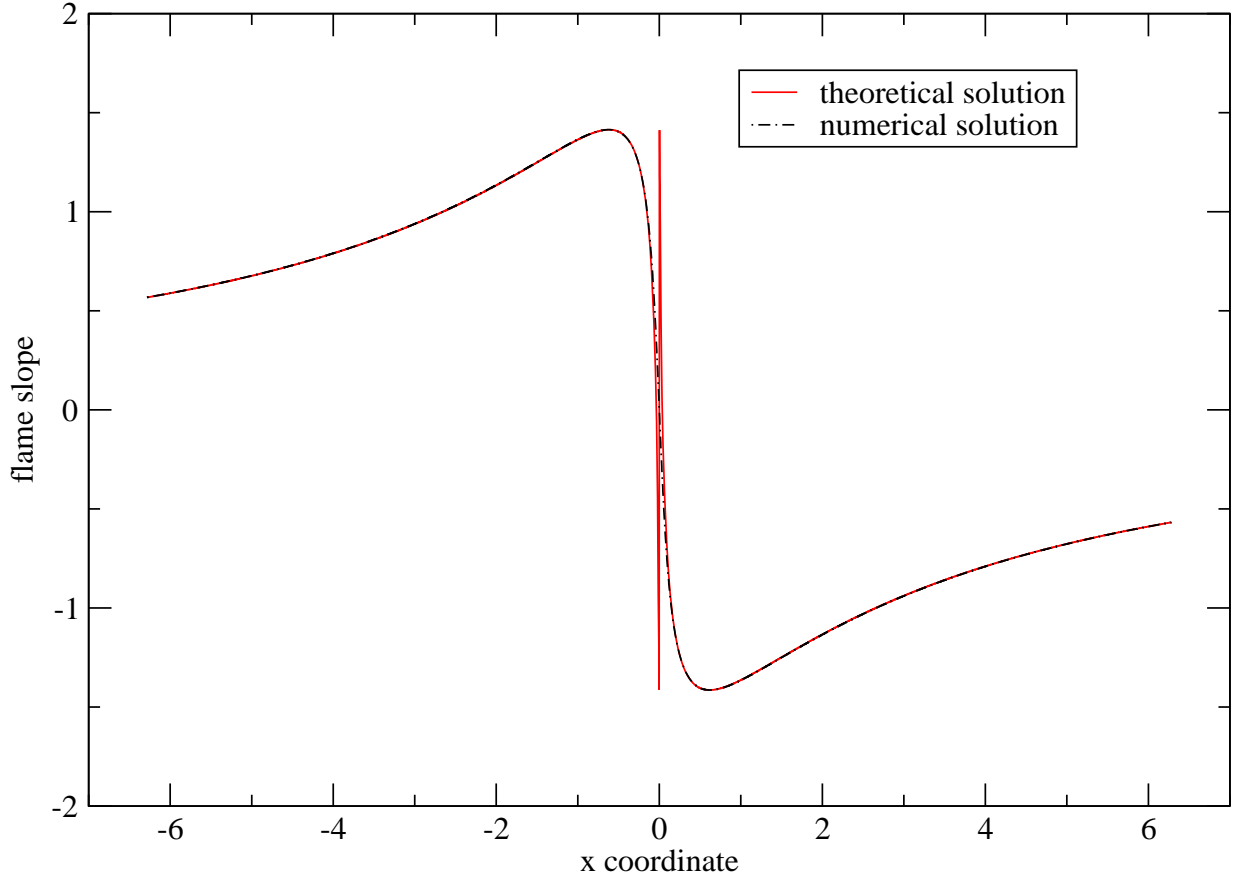

**Figure 5:** Numerical flame-crest slope [from (5)(6)] *vs.* abscissa $x$, and the predicted $\phi_x(x)$ given below Eq.(32); $N = 100$, $\nu = 1/199.5$, $c = +0.5$.

The actual density will vanish for $|B|$ below some $B_{min} > 0$, however. Nice uniform fits of $\phi_x(x)$ (not shown) result from truncating (32) for $|B| < B_c = B_{max} \sin(\theta_c)$, with $\int_0^{B_c} \rho \, dB = 0$ not to alter normalisation: *i.e.*, $\tan(\tfrac{1}{2}\theta_c)/\tan(\tfrac{1}{2}\theta_0) \approx e^{\arctan(m)/m} > 1$ if $B_c \ll B_{max}$. $B_c/B_{max}$ is effectively small ($\approx 2e^{1-\pi^2/\sqrt{c}}$) if $0 < c \ll 1$ but happens to stay so for $c \lesssim 0.6$ ($e^{-\pi^2/\sqrt{0.6}} \approx 3 \, 10^{-6}$). The above $\rho(B)$, $\phi_x$ and $B_{max}$ will hopefully yield correct outer solutions at $|x+iB| \gg B_c$, for $c \to 0^+$. For periodic cells, $\tanh(\tfrac{1}{2}B_c)\coth(\tfrac{1}{2}B_{max}) := \sin(\theta_c)$ has the same, $F$-independent, $\theta_c$ that can also yield good fits.

**6. Open problems**.

This work provides analytical descriptions of steady flame shapes obeying the Zhdanov-Trubnikov equation (1), mostly when its non-local non-linearity is destabilizing ($c < 0$) and the front wrinkles are sharper than MS flames. Yet some problems remain pending.

The density and front-slope obtained for large wrinkles with $c < 0$ are *outer* solutions that hold for $N \to \infty$ only at fixed $c \neq 0$ and $z := x + iB = O(B_{max})$. According to [13], *discrete* pole locations $B_j$ can nevertheless be estimated from $\rho(B)$ by $\int_0^{B_j} \rho(B) dB \approx j - 1/2$: (21) gives $B_j / B_{max} \sim ((j-1/2)/N)^{1/(1-\mu)}$ for $0 < j = O(1)$. Although potentially useful to get the tip curvature $\sim 1/B_1$ when $N$ is moderately large, no analysis of the flame-tip region(s) $O(1/N^{1/(1-\mu)}) \leq |z|/B_{max} \ll 1$ is available.

Although it can yield accurate fits for over-stabilized flames ($c > 0$) up to $c \approx 0.6$, the proposed above density truncation is just an *ad-hoc* stopgap: a bounded-below density support $0 < B_{min} \leq |B| \leq B_{max}$ should instead be introduced from the outset. But alternative methods are needed to handle this, not to mention the flame-tip structure(s) if $1/N$ and $c \gtrless 0$ are small.

A fuller treatment should also describe the 2-tip wrinkles resembling experiments [1] that (1) *often* produces if endowed with Neumann conditions (2 unlike pole-rows per cell [13] [20]).

-----------------------------------